\documentclass{article}
\usepackage{eurosym}
\usepackage{amsfonts}
\usepackage{amsmath}
\usepackage{cite}

\begin{document}

\title{From Duffin-Kemmer-Petiau to Tzou algebras in relativistic wave
equations}
\author{Andrzej Okni\'{n}ski\thanks{
Email: fizao@tu.kielce.pl} \\
Chair of Mathematics and Physics, Politechnika \'{S}wi\c{e}tokrzyska, \\
Al. 1000-lecia PP 7, 25-314 Kielce, Poland}
\maketitle

\begin{abstract}
We study relation between the Duffin-Kemmer-Petiau algebras and some
representations of Tzou algebras. Working in the setting of relativistic
wave equations we reduce, via a similarity transformation, five and ten
dimensional  Duffin-Kemmer-Petiau algebras to three and seven dimensional
Tzou algebras, respectively.
\end{abstract}

\section{Introduction and motivation}
\label{Introduction}

The Duffin-Kemmer-Petiau (DKP) equations \cite%
{Kemmer1939,Duffin1938,Petiau1936} have been becoming increasingly popular
due to their applications to problems in particle and nuclear physics
involving spin $0$ and spin $1$ mesons \cite%
{Yusuk2005,Kasri2008,Chargui2010,Hassanabadi2011,Hamzavi2012,Hassanabadi2012,Boumali2013, Hassanabadi2014,Darroodi2015,Salehi2015,Oliveira2016,Oluwadare2017}%
.

The DKP equations involve $5\times 5$ and $10\times 10$ matrices,
corresponding to representations denoted as $\mathbf{5}$ and $\mathbf{10}$,
respectively, and obeying commutation relations defining the DKP algebra 
\cite{Kemmer1939,Duffin1938,Petiau1936}. However, there are several attempts
to describe spin $0$ and spin $1$ mesons using the so-called Tzou algebras 
\cite{Tzou1957a,Tzou1957b}, see \cite{Beckers1995a,Beckers1995b} as well as 
\cite{Okninski2016} and references therein. Since the spin $0$ as well as
spin $1$ bosons have been described within approaches using different
algebras it seems useful to investigate possible relations between DKP and
Tzou formalisms.

In the next Section Tzou and DKP algebras are described shortly in the
context of relativistic wave equations. In Section \ref{Reduction} the free
DKP equations in momentum representation are converted by a similarity
transformation to equations involving the Tzou algebra, one of them being
the Hagen-Hurley equation. Our results are summarized in the last Section.
In what follows we use notation and conventions described in \cite%
{Okninski2017}.

\section{Relativistic wave equations: Tzou and Duffin-Kemmer-Petiau algebras}
\label{Rwe01}

Equations describing spin $0$ and $1$ bosons, can be written as:%
\begin{equation}
\rho ^{\mu }p_{\mu }\Psi =m\Psi ,  \label{DKP-s0,1}
\end{equation}%
where $p_{\mu }\overset{df}{=}i\frac{\partial }{\partial x^{\mu }}$\ and $%
\rho ^{\mu }$ are matrices with properties described below.

Eq. (\ref{DKP-s0,1}) describes a particle with definite mass if $\rho ^{\mu
} $ obey the Tzou commutation relations \cite%
{Tzou1957a,Tzou1957b,Okninski1981,Beckers1995a,Beckers1995b}:%
\begin{equation}
\sum\nolimits_{\lambda ,\mu ,\nu }^{\lambda }\rho ^{\lambda }\rho ^{\mu
}\rho ^{\nu }=\sum\nolimits_{\lambda ,\mu ,\nu }g^{\lambda \mu }\rho ^{\nu },
\label{Tzou}
\end{equation}%
where we sum over all permutations of $\lambda ,\mu ,\nu $.

A special solution of the Tzou relations (\ref{Tzou}) was constructed in
Ref. \cite{Kemmer1939} in form:%
\begin{equation}
\rho ^{\mu }=\frac{1}{2}\left( \gamma ^{\mu }\otimes I_{4\times
4}+I_{4\times 4}\otimes \gamma ^{\mu }\right) \equiv \beta ^{\mu }.
\label{beta}
\end{equation}%
Such $\beta ^{\mu }$ obey simpler but more restrictive commutation relations 
\cite{Kemmer1939,Duffin1938}:%
\begin{equation}
\beta ^{\lambda }\beta ^{\mu }\beta ^{\nu }+\beta ^{\nu }\beta ^{\mu }\beta
^{\lambda }=g^{\lambda \mu }\beta ^{\nu }+g^{\nu \mu }\beta ^{\lambda },
\label{DuffinKemmer}
\end{equation}%
for which Eq.~(\ref{DKP-s0,1}) leads to the Duffin-Kemmer-Petiau (DKP)
theory of spin $0$ and $1$ mesons, see \cite%
{Kemmer1939,Duffin1938,Petiau1936}. This reducible $16$-dimensional
representation (\ref{beta}) of $\beta ^{\mu }$ matrices (denoted as $\mathbf{%
16}$) can be decomposed as $\mathbf{16}=\mathbf{10}\oplus \mathbf{5}\oplus 
\mathbf{1}$. Representation $\mathbf{10}$ (spin $1$ case) is realized in
terms of $10\times 10$ matrices, while representation $\mathbf{5}$ (spin $0$%
) involves $5\times 5$ matrices, see \cite%
{Kemmer1939,Beckers1995a,Beckers1995b} for explicit formulae of $\beta ^{\mu
}$ (the one-dimensional representation $\mathbf{1}$ is trivial, i.e. all $%
\beta ^{\mu }=0$).

It turns out that in the case of more general Eqs. (\ref{Tzou}) there are
also other representations of $\rho ^{\mu }$ matrices, see \cite%
{Beckers1995a,Beckers1995b} for a review. For example, there are two
representations $\mathbf{7}$\ for which the corresponding $7\times 7$
matrices $\rho ^{\mu }$ yield the Hagen-Hurley equations for spin $1$ bosons 
\cite{HagenHurley1970,Hurley1971,Hurley1974}, see also \cite{Beckers1995a,Beckers1995b} 
and Subsection 2.2 in \cite{Okninski2017}. There are also two sets of $3\times 3$ matrices $\rho
^{\mu }$ obeying (\ref{Tzou}), see \cite{Okninski2003/4}. The aim of this
work is to find possible links between DKP and Tzou representations.

\section{Reduction of Duffin-Kemmer-Petiau representations to Tzou
representations}
\label{Reduction}

Representations $\mathbf{5}$ and $\mathbf{10}$ of the DKP algebra (\ref%
{DuffinKemmer}) are irreducible. However, if we relax these conditions
demanding only Tzou conditions (\ref{Tzou}) then these representations can
be decomposed. We shall work in the momentum representation, $\Psi \left(
x\right) =\psi _{k}e^{-ik\cdot x}$, $\rho ^{\mu }k_{\mu }\psi _{k}=m\psi
_{k} $.

\subsection{Explicit decomposition of representation $\mathbf{5}$}
\label{Representation5}

The $\beta ^{\mu }k_{\mu }$ matrix for irreducible DKP representation $%
\mathbf{5}$ of matrices $\beta ^{\mu }$\ reads:%
\begin{equation}
\beta ^{\mu }k_{\mu }=\left( 
\begin{array}{ccccc}
0 & 0 & 0 & 0 & k_{0} \\ 
0 & 0 & 0 & 0 & k_{1} \\ 
0 & 0 & 0 & 0 & k_{2} \\ 
0 & 0 & 0 & 0 & k_{3} \\ 
k_{0} & -k_{1} & -k_{2} & -k_{3} & 0%
\end{array}%
\right) ,  \label{DKP5}
\end{equation}%
while the matrix $\hat{\rho}^{\mu }k_{\mu }$, corresponding to reducible
Tzou representation $\mathbf{3}\oplus \mathbf{1}\oplus \mathbf{1}$ (note
that $\hat{\rho}^{\mu }$ fulfill conditions (\ref{Tzou})), is:%
\begin{equation}
\hat{\rho}^{\mu }k_{\mu }=\left( 
\begin{array}{ccccc}
0 & 0 & k_{0}+k_{3} & \vline\,\,0 & 0 \\ 
0 & 0 & k_{1}+ik_{2} & \vline\,\,0 & 0 \\ 
k_{0}-k_{3} & -k_{1}+ik_{2} & 0 & \vline\,\,0 & 0 \\ \hline
0 & 0 & 0 & \vline\,\,0 & 0 \\ 
0 & 0 & 0 & \vline\,\,0 & 0%
\end{array}%
\right) .  \label{Tzou5}
\end{equation}%
Both matrices have the same characteristic polynomials, hence there exists a
similarity transformation $V$ converting $\beta ^{\mu }k_{\mu }$ to $\hat{%
\rho}^{\mu }k_{\mu }$, i.e. $V\beta ^{\mu }k_{\mu }V^{-1}=\hat{\rho}^{\mu
}k_{\mu }$. To find $V$ we first diagonalize both matrices $S^{-1}\beta
^{\mu }k_{\mu }S=$ \textrm{diag }$\left( \kappa ,-\kappa ,0,0,0\right)
=T^{-1}\hat{\rho}^{\mu }k_{\mu }T$, $\kappa =\sqrt{%
k_{0}^{2}-k_{1}^{2}-k_{2}^{2}-k_{3}^{2}}$, where columns of $S$ and $T$ are
eigenvectors of $\beta ^{\mu }k_{\mu }$ and $\hat{\rho}^{\mu }k_{\mu }$,
respectively:

\begin{eqnarray}
S &=&\left( 
\begin{array}{ccccc}
\frac{k_{0}}{\kappa } & -\frac{k_{0}}{\kappa } & 1 & 0 & 0 \\ 
\frac{k_{1}}{\kappa } & -\frac{k_{1}}{\kappa } & 0 & 1 & 0 \\ 
\frac{k_{2}}{\kappa } & -\frac{k_{2}}{\kappa } & 0 & 0 & 1 \\ 
\frac{k_{3}}{\kappa } & -\frac{k_{3}}{\kappa } & \frac{k_{0}}{k_{3}} & -%
\frac{k_{1}}{k_{3}} & -\frac{k_{2}}{k_{3}} \\ 
1 & 1 & 0 & 0 & 0%
\end{array}%
\right) ,  \label{S} \\
T &=&\left( 
\begin{array}{ccccc}
\frac{k_{0}+k_{3}}{k_{1}+ik_{2}} & \frac{k_{0}+k_{3}}{k_{1}+ik_{2}} & \frac{%
k_{1}-ik_{2}}{k_{0}-k_{3}} & 0 & 0 \\ 
1 & 1 & 1 & 0 & 0 \\ 
\kappa \frac{k_{1}-ik_{2}}{k_{1}^{2}+k_{2}^{2}} & \kappa \frac{-k_{1}+ik_{2}%
}{k_{1}^{2}+k_{2}^{2}} & 0 & 0 & 0 \\ 
0 & 0 & 0 & 1 & 0 \\ 
0 & 0 & 0 & 0 & 1%
\end{array}%
\right) .  \label{T}
\end{eqnarray}%
Finally, we have $V=TS^{-1}$. We have assumed that denominators in Eqs. (\ref%
{S}), (\ref{T}) do not vanish. However, in special cases when some (or all)
denominators do vanish (i.e. if $k_{3}=0$, or $k_{1}=k_{2}=0$, or $%
k_{1}=k_{2}=k_{3}=0$), transformations $S$, $T$, $V$ still exist.

Note that $V$ depends on $k_{\mu }$ and thus $V\beta ^{\mu }V^{-1}\neq \hat{%
\rho}^{\mu }$. Accordingly, while matrices $\beta ^{\mu }$ belong to
representation $\mathbf{5}$ of the DKP algebra (they are also a
representation of the Tzou algebra), matrices $\hat{\rho}^{\mu }$ are
representation of the Tzou algebra only. It follows from Eq. (\ref{Tzou5})
that the irreducible representation $\mathbf{3}$ of the Tzou algebra is:%
\begin{equation}
\rho ^{\mu }k_{\mu }=\left( 
\begin{array}{ccc}
0 & 0 & k_{0}+k_{3} \\ 
0 & 0 & k_{1}+ik_{2} \\ 
k_{0}-k_{3} & -k_{1}+ik_{2} & 0%
\end{array}%
\right) ,  \label{3}
\end{equation}%
see also Eqs. (14), (16) in \cite{Okninski2003/4}.

\subsection{Decomposition of representation $\mathbf{10}$}
\label{Representation10}

Matrices $\beta ^{\mu }$ of the DKP representation $\mathbf{10}$ can be
found in \cite{Kemmer1939,Beckers1995a,Beckers1995b}. The matrix $\beta
^{\mu }k_{\mu }$ reads:\medskip 
\begin{equation}
\beta ^{\mu }k_{\mu }=\left( 
\begin{array}{cccccccccc}
0 & 0 & 0 & 0 & 0 & 0 & k_{1} & k_{0} & 0 & 0 \\ 
0 & 0 & 0 & 0 & 0 & 0 & k_{2} & 0 & k_{0} & 0 \\ 
0 & 0 & 0 & 0 & 0 & 0 & k_{3} & 0 & 0 & k_{0} \\ 
0 & 0 & 0 & 0 & 0 & 0 & 0 & 0 & k_{3} & -k_{2} \\ 
0 & 0 & 0 & 0 & 0 & 0 & 0 & -k_{3} & 0 & k_{1} \\ 
0 & 0 & 0 & 0 & 0 & 0 & 0 & k_{2} & -k_{1} & 0 \\ 
-k_{1} & -k_{2} & -k_{3} & 0 & 0 & 0 & 0 & 0 & 0 & 0 \\ 
k_{0} & 0 & 0 & 0 & k_{3} & -k_{2} & 0 & 0 & 0 & 0 \\ 
0 & k_{0} & 0 & -k_{3} & 0 & k_{1} & 0 & 0 & 0 & 0 \\ 
0 & 0 & k_{0} & k_{2} & -k_{1} & 0 & 0 & 0 & 0 & 0%
\end{array}%
\right) \hspace{-0.35cm}  \label{DKP10}
\end{equation}%
\medskip and the matrix $\hat{\rho}^{\mu }k_{\mu }$, with matrices $\hat{\rho%
}^{\mu }$\ corresponding to reducible Tzou representation $\mathbf{7}\oplus 
\mathbf{1}\oplus \mathbf{1}\oplus \mathbf{1}$ (note again that $\hat{\rho}%
^{\mu }$ fulfill conditions (\ref{Tzou})), is:\medskip 
\begin{equation}
\hat{\rho}^{\mu }k_{\mu }=\left( 
\begin{array}{cccccccccc}
0 & 0 & 0 & -ik_{1} & -ik_{0} & -k_{3} & k_{2} & \vline\,\,0 & 0 & 0 \\ 
0 & 0 & 0 & -ik_{2} & k_{3} & -ik_{0} & -k_{1} & \vline\,\,0 & 0 & 0 \\ 
0 & 0 & 0 & -ik_{3} & -k_{2} & k_{1} & -ik_{0} & \vline\,\,0 & 0 & 0 \\ 
-ik_{1} & -ik_{2} & -ik_{3} & 0 & 0 & 0 & 0 & \vline\,\,0 & 0 & 0 \\ 
ik_{0} & -k_{3} & k_{2} & 0 & 0 & 0 & 0 & \vline\,\,0 & 0 & 0 \\ 
k_{3} & ik_{0} & -k_{1} & 0 & 0 & 0 & 0 & \vline\,\,0 & 0 & 0 \\ 
-k_{2} & k_{1} & ik_{0} & 0 & 0 & 0 & 0 & \vline\,\,0 & 0 & 0 \\ \hline
0 & 0 & 0 & 0 & 0 & 0 & 0 & \vline\,\,0 & 0 & 0 \\ 
0 & 0 & 0 & 0 & 0 & 0 & 0 & \vline\,\,0 & 0 & 0 \\ 
0 & 0 & 0 & 0 & 0 & 0 & 0 & \vline\,\,0 & 0 & 0%
\end{array}%
\right)   \label{Tzou10}
\end{equation}

\medskip Proceeding exactly as in Subsection \ref{Representation5} we
construct matrix $V$ of the similarity transformation $V\beta ^{\mu }k_{\mu
}V^{-1}=\hat{\rho}^{\mu }k_{\mu }$. Finally, the matrix $\rho ^{\mu }k_{\mu }
$ is the $7\times 7$ non-zero block in Eq. (\ref{Tzou10}). This corresponds
to decomposition of representation $\mathbf{10}$ as $\mathbf{7}\oplus 
\mathbf{1}\oplus \mathbf{1}\oplus \mathbf{1}$. Equation $\rho ^{\mu }p_{\mu
}\Psi =m\Psi $ is one of the Hagen-Hurley equations describing spin $1$
bosons, cf. Eq. (16) and the text below in \cite{Okninski2017}.

\section{Summary}
\label{Summary}

We have demonstrated that the free Duffin-Kemmer-Petiau equations in the
momentum representation can be converted by application of a similarity
transformation to equations involving Tzou algebra. More exactly, there
exists a matrix $V$, dependent on $k_{\mu }$, such that: 
\begin{subequations}
\label{V}
\begin{align}
V\left( \beta ^{\mu }k_{\mu }-m\right) V^{-1}& =\hat{\rho}^{\mu }k_{\mu }-m,
\label{V1} \\
V\beta ^{\mu }V^{-1}& \neq \hat{\rho}^{\mu },  \label{V2}
\end{align}
\end{subequations}
where $\beta ^{\mu }$ fulfill the Duffin-Kemmer-Petiau relations (\ref%
{DuffinKemmer}), while $\hat{\rho}^{\mu }$ belong to the (reducible) Tzou
algebra (\ref{Tzou}). Therefore, in the case of spin $1$ bosons we have
constructed a link (a similarity transformation) between the
Duffin-Kemmer-Petiau equation with $10\times 10$ matrices $\beta ^{\mu }$
and Hagen-Hurley equation involving $7\times 7$ matrices $\rho ^{\mu }$
belonging to Tzou algebra. Analogously, in the case of spin $0$ bosons there
is a similar link between the Duffin-Kemmer-Petiau equation with $5\times 5$
matrices $\beta ^{\mu }$ and equation involving $3\times 3$ matrices $\rho
^{\mu }$ belonging to Tzou algebra.

\end{document}